# A quantum circuit simulator and its applications on Sunway TaihuLight supercomputer

Zhimin Wang[1], Zhaoyun Chen[2,3], Shengbin Wang[1], Wendong Li[1], Yongjian Gu[1*], Guoping Guo[2,3*], Zhiqiang Wei[1,4*]

[1] College of Information Science and Engineering, Ocean University of China, Qingdao 266100, China
[2] CAS Key Laboratory of Quantum Information, University of Science and Technology of China, Hefei 230026, China
[3] Origin Quantum Computing Company Limited, Hefei 230026, China
[4] High Performance Computing Center, Pilot National Laboratory for Marine Science and Technology (Qingdao), Qingdao 266100, China
[*] Correspondence author: yjgu@ouc.edu.cn (Y. Gu); gpguo@ustc.edu.cn (G. Guo); weizhiqiang@ouc.edu.cn (Z. Wei)

**Abstract** Classical simulation of quantum computation is vital for verifying quantum devices and assessing quantum algorithms. We present a new quantum circuit simulator developed on the Sunway TaihuLight supercomputer. Compared with other simulators, the present one is distinguished in two aspects. First, our simulator is more versatile. The simulator consists of three mutually independent parts to compute the full, partial and single amplitudes of a quantum state with different methods. It has the function of emulating the effect of noise and support more kinds of quantum operations. Second, our simulator is of high efficiency. The simulator is designed in a two-level parallel structure to be implemented efficiently on the distributed many-core Sunway TaihuLight supercomputer. Random quantum circuits can be simulated with 40, 75 and 200 qubits on the full, partial and single amplitude, respectively. As illustrative applications of the simulator, we present a quantum fast Poisson solver and an algorithm for quantum arithmetic of evaluating transcendental functions. Our simulator is expected to have broader applications in developing quantum algorithms in various fields.

**Keywords** Quantum computing, Quantum circuit simulator, Sunway TaihuLight, Quantum algorithm

## 1 Introduction

In recent years, tremendous technological progress has been made in the construction of quantum computers, especially with superconducting qubits [1,2]. As these nascent quantum computers become competitive against classical computers in simulating general quantum circuits, an interesting race come to the climax. The quantum beings are eager to accomplish the first demonstration of quantum supremacy [1,3], while classical beings try to push back the classical simulation barrier as far as possible [4-7].

During the race, many novel methods and programs are developed to simulate quantum circuits on classical computers efficiently [8-16]. In fact, classical simulation of quantum computation is vital both for the verification of quantum computers and for the assessment of the correctness and performance of new quantum algorithms. The fundamental task of such simulation is to calculate all or a certain number of amplitudes of quantum states produced by a quantum circuit.

However, it is extremely expensive to simulate quantum computation classically because of the curse of dimensionality, i.e., the memory and time requirements grow exponentially with the number of qubits. For instance, to accurately simulate a quantum system with 50 qubits, one needs a classical computer with slightly more than 16 Petabytes of memory (with double precision). Moreover, increasing the number of qubits by one requires a doubling of the amount of memory space. Performing such a large-scale computation requires one to take advantage of the state-of-the-art high-performance distributed computation.

In the present work, we develop a new quantum circuit simulator on the Sunway TaihuLight supercomputer. Albeit other simulators have been developed on supercomputers including Sunway TaihuLight [11,13,16], our simulator is designed to be a powerful tool for quantum algorithm research. The simulator consists of three mutually independent sub-programs to calculate the full, partial and single amplitudes of a quantum state with three completely different methods. Therefore, a wide range of number of qubits and circuit depths can be covered. This could provide choices when people execute quantum algorithms of different fields. In addition, it can emulate the effect of noise and support more quantum operations, such as the controlled and inverse operations on a group of gates, which are very useful in practical applications. On the other hand, the efficiency of the simulator is high. The algorithms of the simulator has a two-level parallel structure to fully take advantage of the Sunway system architecture. We can simulate random quantum circuits with 40, 75 and 200 qubits on the full, partial and single amplitude, respectively. With this simulator, we further develop quantum algorithms for solving the Poisson equations and for quantum arithmetic of evaluating transcendental functions.

## 2 Methods

The present quantum circuit simulator consists of three mutually independent sub-programs, referred to as three working modes of the simulator, i.e. full amplitude, partial amplitude and single amplitude mode. The fundamental methodologies for the three modes are completely different. They are, respectively, direct evolution of quantum state, circuit partition by decomposing controlled-Z gate [10], and the complex undirected graphical model [9]. In addition, noisy one- and two-qubit gates are defined to emulate the effect of noise. A description of the instruction set of our simulator as well as an illustrative example of the input and output are given in the Appendix A.

2.1 Sunway TaihuLight Supercomputer

Before proceeding to the details of the simulation techniques, we first give a brief introduction of the classical hardware. Our simulator is developed based on the Sunway

TaihuLight at the National Supercomputer Center in Wuxi, China. The Sunway TaihuLight is so far the most powerful supercomputer in China. It can reach a peak performance of 125 PFlops, and had ranked the first in the TOP500 list for four times in the years of 2016 and 2017.

The supercomputer consists of 40960 homegrown processors called SW26010. Each SW26010 processor contains four core-groups. Each core-group contains one management processing element (MPE, hereafter called master core) with a memory space of 8GB, and 64 computing processing elements (CPE, hereafter called slave core) in an 8×8 array [17]. The master core supports the programming languages of C, C++ and FORTRAN, while the slave core supports C and FORTRAN. Within a core-group, the 64 slave cores can communicate with each other in a few cycles.

In the present work, one core-group is set as a unique MPI process. For simplicity, when mentioning a computational node, it refers to one core-group, namely 1 master core plus 64 slave cores. The simulator is written by C++ language.

To take full advantage of the system architecture of Sunway TaihuLight, we implement algorithms of the three working modes in a two-level parallel way. More specifically, the entire simulation are first divided equally to the available nodes, which is the first level of parallel. In each node corresponding to a unique MPI process, the computing task is further assigned to the 64 slave cores equally, while the master core is responsible for the process control and I/O operation. This is the second level of parallel. The specific designs of algorithms are discussed in the subsequent sections.

2.2 Full amplitude mode

The full amplitude mode of the simulator is an instance of the so-called Schrodinger simulation. It is based on the direct evolution of quantum state. All the information of the quantum state is precisely maintained and updated step-by-step throughout the simulation. The Schrodinger approach is straightforward, and it could provide a great speed in simulating low-width circuits. However, when processing many-qubits circuits, it requires a significant amount of RAM to store all amplitudes. In the present work, we use at most 16384 computational nodes, roughly 10% of the computing resource of Sunway TaihuLight, and can simulate a quantum circuit with up to 40 qubits on this mode.

Now we use the single-qubit and controlled two-qubit operations as examples to illustrate the distributed implementation of Schrodinger simulation. It is well known that an *n*-qubit quantum state can be represented by Dirac notations and column vectors as follows,

$$
\begin{aligned}
|\varphi\rangle &= \sum_{i=0}^{2^n-1} \alpha_i |i\rangle = \sum_{i_{n-1}\cdots i_0 = \{0,1\}^n} \alpha_i |i_{n-1}\cdots i_k \cdots i_0\rangle \quad (Dirac) \\
&= (\alpha_0, \alpha_1, \cdots, \alpha_{2^n-1})^T \quad (column\ vector)
\end{aligned}
\qquad (1)
$$

where the decimal and binary index are related by $i = i_{n-1} \times 2^{n-1} + \cdots + i_k \times 2^k + \cdots + i_0 \times 2^0$. In practice, we store all the amplitudes $\alpha_i$ in the memory during the simulation and update them according to the action of unitary operations.

Let $U^k$ represents a single-qubit gate acting on the $k^{\text{th}}$ qubit, namely $i_k$ in Eq. (1). It can be easily verified that the amplitudes can be updated in the following way,

$$\begin{pmatrix} \alpha'_i \\ \alpha'_{i+2^k} \end{pmatrix} = U^k \begin{pmatrix} \alpha_i \\ \alpha_{i+2^k} \end{pmatrix} = \begin{pmatrix} a & b \\ c & d \end{pmatrix} \begin{pmatrix} \alpha_i \\ \alpha_{i+2^k} \end{pmatrix} = \begin{pmatrix} a\alpha_i + b\alpha_{i+2^k} \\ c\alpha_i + d\alpha_{i+2^k} \end{pmatrix}. \quad (2)$$

for all $(i)_{10} = (i_{n-1} \cdots i_k \cdots i_0)_2$ with $i_k = 0$

Note that the amplitudes indexed by $i_k=1$ are calculated when traversing the index of $i+2^k$. As can be seen from the equation, for one action of $U^k$, all the $2^n$ amplitudes are changed. Thus, the one single-qubit operation corresponds to a computation scale of $2^n$ additions and multiplications. The controlled two-qubit operation can be implemented similarly. Let $CU^{q,k}$ represents a controlled two-qubit gate, where the qubit $q$ ($k$) is the control (target) bit. That is to say, when $i_q$ is zero, the gate $CU^{q,k}$ will do nothing; when $i_q$ is 1, the gate performs the same transformation as Eq. (2). This can be formalized as

$$\begin{cases} \alpha'_i = \alpha_i, & \text{with } i_q = 0; \\ \begin{pmatrix} \alpha'_i \\ \alpha'_{i+2^k} \end{pmatrix} = \begin{pmatrix} a\alpha_i + b\alpha_{i+2^k} \\ c\alpha_i + d\alpha_{i+2^k} \end{pmatrix}, & \text{with } i_q = 1, i_k = 0 \end{cases}. \quad (3)$$

for all $(i)_{10} = (i_{n-1} \cdots i_q \cdots i_k \cdots i_0)_2$ or $(i_{n-1} \cdots i_k \cdots i_q \cdots i_0)_2$

Here we remark that although the above single-qubit and controlled two-qubit operations are enough as they form a universal set for quantum computation [18], our simulator could support more quantum gates and operations. They are very useful in the practical design of quantum circuits. Particularly, the simulator supports arbitrary single-quit rotation gates, controlled operation on a group of gates, and inverse operation on a group of gates, etc. (see Appendix A for details).

The above two equations are of great importance because they lend the process of updating amplitudes to parallelization and distribution. That is, they update amplitudes via $2^n$ computations of $a\alpha_i+b\alpha_j$ as shown in Eq. (3), not by multiplying a full $2^n \times 2^n$ matrix on the column vector. Such equations can be implemented in a way of two-level parallel. Specially, all the amplitudes are divided equally to the nodes and stored in the corresponding master cores. Then the master core in each node calls the slave cores to update the amplitudes in parallel.

In summary, the program of this mode proceeds in the following three steps:

1$^{\text{st}}$. Configure the computation nodes. Then every node parses the script to obtain a linked list recording instructions of the quantum circuit.

2$^{\text{nd}}$. Assign all the amplitudes equally to the nodes. The amplitudes are initialized as zero in the master core of each node.

3$^{\text{rd}}$. The master core traverses every node of the linked list in turn, and prepares the computing parameters, including the matrix coefficients of the gate, the number of amplitudes of each node, starting address of the target amplitude, etc. Then the master core assign the task of updating the amplitudes equally to the 64 slave cores. The slave cores get the requited data using the address information according to Eqs. (2) or (3), and compute the new amplitude values, and then sent them back to the same position in the master core.

2.2 Partial amplitude mode

The partial amplitude mode use a hybrid algorithm to simulate a quantum circuit with more than 50 qubits but of limited depth. Generally, in this mode the original quantum circuit are divided into several sub-circuits with less qubits, which are then simulated independently using the same method as the full amplitude mode. With 16384 computational nodes, we can simulate a quantum circuit with up to 75 qubits under this mode. Below is a brief introduction of the partition scheme of the circuit. More information can be found in our previous paper [10].

The controlled-$Z$ gates can be decomposed into the projection and single-qubit $Z$ gates as follows,

$$CZ^{i,j} = P_0^i \otimes I^j + P_1^i \otimes Z^j = \begin{pmatrix} 1 & 0 \\ 0 & 0 \end{pmatrix}^i \otimes I^j + \begin{pmatrix} 0 & 0 \\ 0 & 1 \end{pmatrix}^i \otimes Z^j. \qquad (4)$$

The superscripts represent that qubit $i$ is the control qubit and qubit $j$ the target qubit. On the left hand side of the equation, qubits $i$ and $j$ are entangled, while on the right hand side they are independent. Therefore, after decomposing the $CZ$ gate, the quantum states of qubits $i$ and $j$ can evolve independently, and then be recombined to get the final state. This turns out to be a very useful method of reducing the memory requirements when simulating a quantum circuit with many qubits.

Now we take a quantum circuit with 8 qubits and 8 layers of depth as an example to illustrate the partition scheme. The circuit is shown in Fig. 1. The circuit is made up of two blocks, that is, the upper block with qubits from 0 to 3, and the lower one with the other qubits. The two blocks are entangled by the $CZ$ gates in $7^{th}$ and $8^{th}$ layer. The entanglement between the two blocks can be dismissed by decomposing the two $CZ$ gates in turn, as shown in Fig. 1. After the decomposition, the original circuit results in four circuits, whose upper and lower blocks are untangled. Then each of the four circuits can be divided into two sub-circuits with a half number of qubits, which can be simulated independently. Therefore, the task of simulating the original circuit with 8 qubits is converted to the simulation of 8 independent sub-circuits with 4 qubits. The number of amplitudes stored in the memory is reduced from $2^8$ to $2^7$. Since the sub-circuits are simulated in a parallel way, the time span of the simulation is also reduced.

There are also restrictions on the partition scheme. The gates crossing the dividing line should be the controlled two-qubit gate, such as the $CNOT$ and $CZ$ gates, not the gate like SWAP. Furthermore, the number of sub-circuits grows exponentially with the number of decomposed $CZ$ gates. For example, if there is one more $CZ$ gate crossing the dividing line between qubits 3 and 4 in Fig. 1, the partition is not efficient. Therefore, this method is suitable for quantum circuits with low depth and large sampling number (the large sampling number is originate from the fact that all the sub-circuits are simulated on the full amplitude mode).

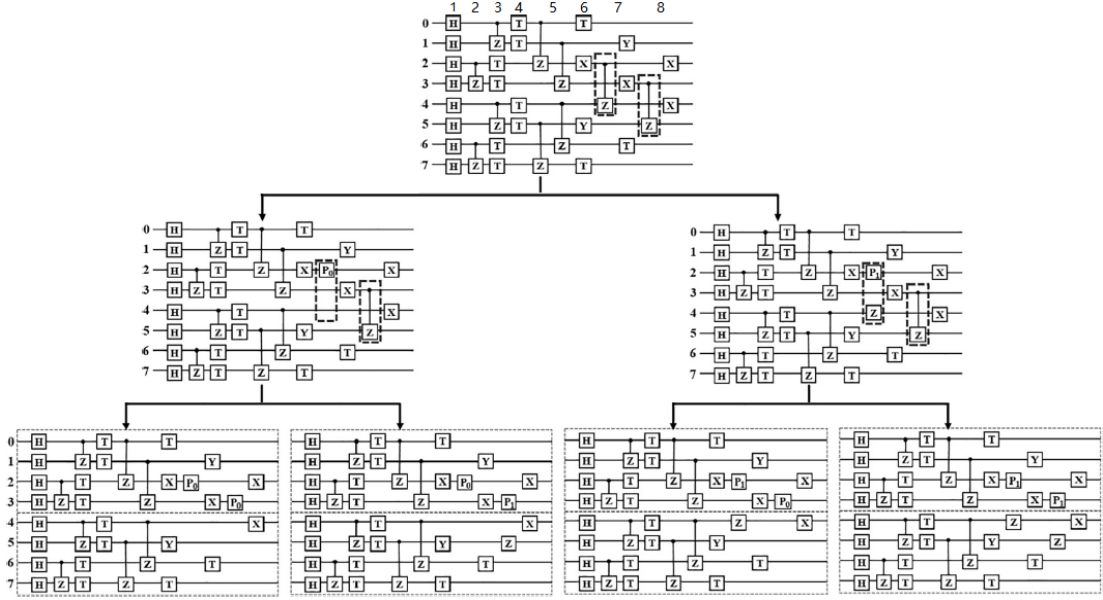

**Fig. 1** The partition scheme for a quantum circuit with 8 qubits and 8 depths [10]. The first line shows the original circuit. In the second line, the *CZ* gate in $7^{th}$ depth is decomposed to projection and single-qubit *Z* gate, then the original circuit converts to two circuits. In the third line, the *CZ* gate in $8^{th}$ depth is further decomposed, and the number of circuit is again double. In each of the final 4 circuits, the upper and lower parts is untangled, and they can be simulated independently

In summary, the program for the partial amplitude mode proceeds in the following four steps:

$1^{st}$. Configure the computation nodes. Then every node parses the script to extract the gates. Judge whether the gates crossing the dividing line is the controlled two-qubit gates, and decompose it by doubling the circuit. The dividing line is always set to be in the middle of qubits.

$2^{nd}$. Cut each of the final circuits into two sub-circuits along the dividing line. There should be $2^{c+1}$ sub-circuits generated, where *c* is the number of decomposed gates. Establish a linked list of quantum gates for each sub-circuit.

$3^{rd}$. Assign the task of simulating the sub-circuits equally to the nodes. The result of assignment would be that one node simulates one sub-circuit, one node simulates several sub-circuits, or several nodes simulate one sub-circuit. The simulations are implemented in the same way as the full amplitude mode.

$4^{th}$. Combine the state of each sub-circuit to get the final states.

2.3 Single amplitude mode

The single amplitude mode makes use of undirected graphical model to be capable of simulating quantum circuit with much more qubits. Broadly, the original quantum circuit is first mapped to an undirected graphical model, then the undirected graph is split into several ones by fixing the value of variables, and then the resulting graphs are processed in parallel by the vertical variable elimination algorithm.

The undirected graph model is a way of interpreting the relation between the change of bit values of qubit state and the quantum gates. Naturally, the bit of state will change with actions of a sequence of quantum gates. We define a sequence of Boolean variables to describe the change. For example, being acted upon by the Pauli-X and H gate in sequence, the state $|0\rangle$ will be first changed to $|1\rangle$, then to $1/\sqrt{2}(|0\rangle-|1\rangle)$. Then the corresponding Boolean variables are $a_0=0$, $a_1=1$, and $a_2=\{0, 1\}$, respectively. The undirected graph is constructed based on the Boolean variables and quantum gates. Specially, each Boolean variable in the circuit corresponds exactly to one vertex in the graph, and one or multiple gates in the circuit result in one edge in the graph.

The rule of mapping a quantum circuit to an undirected graph is simple and easy to follow [9]. It is summarized to four cases as shown in Fig. 2. For the diagonal one- or two-qubit gate, it does not change the Boolean variable, so the vertices corresponding to the same variable merge into one. For example, the CZ gate will transform the state $|11\rangle$ to $-|11\rangle$ without flipping of the bit, so the input and output vertices are merged as shown in Fig. 2(c). The cross lines in the graph should be considered as one line, which corresponds to one gate, as shown in Fig. 2(d). Fig. 3 presents an example to further illustrate the mapping of a circuit to undirected graph.

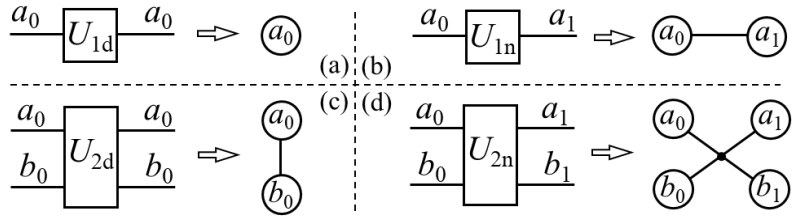

**Fig. 2** The quantum circuit representations of qubits and gates, and the corresponding undirected graphical model. (a) The diagonal one-qubit gate, (b) the non-diagonal one-qubit gate, (c) the diagonal two-qubit gate, (d) the non-diagonal two-qubit gate. The Boolean variables of $a_0$, $a_1$, $b_0$ and $b_1$ represent the bit value of the state, which is 0 or 1. The vertices in the graph corresponds to the Boolean variables in the circuit, and the edges corresponds to the gates

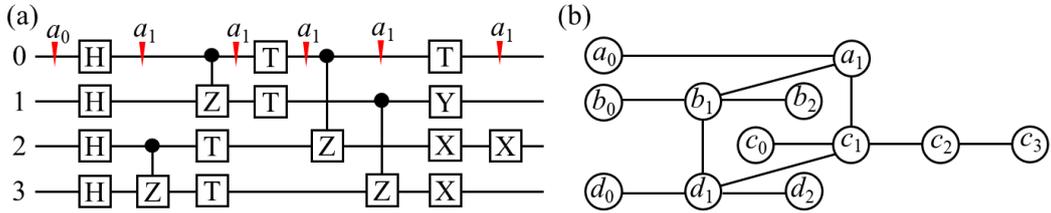

**Fig. 3** An illustration example of mapping a quantum circuit to the undirected graphical model. (a) A quantum circuit adapted from the part of Fig. 1, (b) the corresponding undirected graph. The triangle symbols are used to explain the change of Boolean variables along the worldline of qubit 0. Note that since the cross lines are considered as one line in the undirected graph, the vertices are rearranged to avoid the false crossover

After getting the undirected graph, tensor techniques are used to process it. One edge in the graph corresponds to a particular tensor, and the number of vertices connecting to the edge is the rank of the tensor. For example, the edge in Fig. 2(d) corresponds to a tensor $T$ of rank 4, with $2^4$ elements indexed by $T_{a_0 b_0 a_1 b_1}$. The elements of tensor $T$ are filled using $U_{2n}$ in the lexicographical order of the index, such as that $T_{00,00}=(U_{2n})_{0,0}$, $T_{00,10}=(U_{2n})_{0,2}$, $T_{01,00}=(U_{2n})_{1,0}$, $T_{10,00}=(U_{2n})_{2,0}$ and so on.

There are two kinds of processes performed on the undirected graph, which are edge merging and vertex elimination. Edge merging means that two edges connecting to the same vertex are merged to one. This is actually to merge two tensors with the same subscript into one. For instance, suppose that the edge between vertexes $b_0$ and $b_1$ in Fig. 3(b) corresponds to a tensor $A_{b_0 b_1}$, and the edge between vertex $b_1$ and $d_1$ corresponds to $B_{b_1 d_1}$, then the two edges merges into one to get a higher-rank tensor as

$$C_{b_0 b_1 d_1} = A_{b_0 b_1} B_{b_1 d_1}.$$

Vertex elimination reduces the number of vertexes connecting to a particular edge. This is actually a variant of tensor contraction. We do this using two different methods, of which one is a differential way and the other an integral way. In the differential method, the variable corresponding to a vertex is fixed to be 0 and 1 [19]. For example, the vertex $b_1$ in Fig. 3(b) is fixed to 0, then the tensor $B_{b_1 d_1}$ is converted to $B_{0 d_1}$. Thus, the tensor rank is reduced from 2 to 1, and the number of elements from $2^2$ to 2. The expense of this method is that it doubles the graph. That is, the graph needs to be computed twice with the target variable being 0 and 1, respectively. In the integral method, all the elements of a tensor corresponding to a specific subscript are summed over to eliminate that index. For instance, the subscript $b_1$ in tensor $C_{b_0 b_1 d_1}$ is eliminated by

$$C'_{b_0 d_1} = C_{b_0 0 d_1} + C_{b_0 1 d_1},$$ so the vertex $b_1$ is eliminated from the edge corresponding to tensor $C_{b_0 b_1 d_1}$.

In summary, the program for the single amplitude mode proceeds in the following four steps:

1st. Configure the computation nodes. Then every node parses the script to obtain a linked list recording instructions of the quantum circuit. Map the quantum circuit to the undirected graphical model using the linked list.

2nd. Eliminate the vertices in the first and last depth of the graph according to the specified initial and measurement states using the differential vertex elimination method. Since the initial and measurement states are certain, this step does not double the number of graphs.

3rd. Find the top $N$ vertices with the largest number of connecting edges. Then perform the differential vertex elimination on the $N$ vertices, and this result into $2^N$ graphs. Assign the task of simulating the $2^N$ graphs equally to the nodes. (Note that eliminating the top $N$ high-degree vertices would be not the best way of simplifying the graph. The treewidth of the graph really matters, but it is NP-complete hard to determine [9,19]. For simplicity, we choose the top $N$ high-degree vertices to remove at this step.)

4th. For each graph, eliminate all the vertices. Specifically, for each vertex, first merge all the connecting edges into one in the order of rank, and then eliminate this vertex using the integral method. Multiply the elements of the tensors corresponding to the left edges, and obtain the amplitude of each graph. Sum over the amplitude of each graph to get the final amplitude of the state to measure.

2.4 Simulation of the effect of noise

In practical quantum devices, qubits are performed imperfectly. Various kinds of noise would randomly induce errors on the states of qubits. Particularly, in the coming NISQ era, quantum computers have noisy gates unprotected by quantum error correction [20]. Thus, it is important to characterize the effect of noise by classical simulations.

The effect of noise can be described by a series of super operators $\{K_1, K_2, ..., K_s\}$, which satisfy the relation $\sum_i K_i^\dagger K_i = I$. For the single-qubit gate, we consider the following six kinds of noise,

$$
\begin{aligned}
\text{Bit flip}: & \quad K_1 = \sqrt{p}\begin{bmatrix}1 & 0\\ 0 & 1\end{bmatrix},\ K_2 = \sqrt{1-p}\begin{bmatrix}0 & 1\\ 1 & 0\end{bmatrix};\\
\text{Phase flip}: & \quad K_1 = \sqrt{p}\begin{bmatrix}1 & 0\\ 0 & 1\end{bmatrix},\ K_2 = \sqrt{1-p}\begin{bmatrix}1 & 0\\ 0 & -1\end{bmatrix};\\
\text{Bit-Phase flip}: & \quad K_1 = \sqrt{p}\begin{bmatrix}1 & 0\\ 0 & 1\end{bmatrix},\ K_2 = \sqrt{1-p}\begin{bmatrix}0 & -i\\ i & 0\end{bmatrix};\\
\text{Amplitude Damping}: & \quad K_1 = \begin{bmatrix}1 & 0\\ 0 & \sqrt{1-p}\end{bmatrix},\ K_2 = \begin{bmatrix}0 & \sqrt{p}\\ 0 & 0\end{bmatrix};\\
\text{Phase Damping}: & \quad K_1 = \begin{bmatrix}1 & 0\\ 0 & \sqrt{1-p}\end{bmatrix},\ K_2 = \begin{bmatrix}0 & 0\\ 0 & \sqrt{p}\end{bmatrix};\\
\text{Depolarizing}: & \quad K_1 = \sqrt{1-3p/4}\begin{bmatrix}1 & 0\\ 0 & 1\end{bmatrix},\ K_2 = \sqrt{p}/2\begin{bmatrix}0 & 1\\ 1 & 0\end{bmatrix},\ K_3 = \sqrt{p}/2\begin{bmatrix}0 & -i\\ i & 0\end{bmatrix},\ K_4 = \sqrt{p}/2\begin{bmatrix}1 & 0\\ 0 & -1\end{bmatrix}.
\end{aligned}
\tag{5}
$$

The value $p$ in the equation is on $[0, 1]$, which is proportional to the noise intensity. Specifically, for the first three kinds of noise, when $p$ approaches 1, the noise close to zero; for the last three kinds of noise, when $p$ approaches zero, the noise close to zero.

For the two-qubit gate, the noise operators are defined as the Kronecker products of single-qubit gates. For example, suppose the noise operators of single-qubit gates are $\{K_1, K_2\}$ and $\{M_1, M_2\}$, respectively. Then, the noise operators of two-qubit gate are $\{K_1 \otimes M_1, K_1 \otimes M_2, K_2 \otimes M_1, K_2 \otimes M_2\}$.

In the program, the procedure of simulating the noise goes as follows:

1st. Determine the class of quantum gates specified to be noisy and the kind of noise. Let every operator of $\{K_1, K_2, ..., K_s\}$ act on the present quantum state using the same method as the full amplitude mode. Then calculate the modulus of the states, namely the probabilities of the states.

2nd. Produce a random number between 0 and 1, and compare it with the above sequence of probabilities, then determine which sub-operator $K_i$ to be used. Multiply the matrix $K_i$ with the quantum gate to obtain a new matrix, i.e., the noisy gate.

3$^{rd}$. Update the state by the new matrix using the same method as the full amplitude mode. Finally, normalize the quantum state (the noisy gate may not be unitary).

To sum up, we have discussed the basic principles of the full, partial and single amplitude modes, as well as the way of defining noisy gate to emulate the effect of noise. Subsequently, we introduce numerical results and applications of the present simulator.

## 3 Results and discussion

To characterize the performance of the simulator, we first implement the random quantum circuits (RQCs) generated using the prescription of Google [21]. Then we demonstrate the quantum circuits for solving the Poisson equations and for the quantum arithmetic of evaluating transcendental functions. Here, we remark that the quantum fast Poisson solver and quantum arithmetic algorithms are implemented mainly on the full amplitude mode since these circuits have relatively few qubits and high depth. We leave such applications to future work that the partial and single amplitude modes as well as the function of emulating the effect of noise are exploited.

### 3.1 Implementation of RQCs

The full amplitude mode is the foundation of the other two modes, because the resulting sub-circuits in partial and single modes are finally simulated using the same method as the full amplitude mode. The main factor of limiting the computing speed of full amplitude mode is the data communication between nodes. According to Eqs. (2) and (3), when updating one term of amplitudes $\alpha_i$, one need another term $\alpha_{i+2^k}$, which may be stored in another core-group or another SW26010 processors. As shown in Fig. 4, for the one-qubit gate, the speed of computation on a state stored in one core-group (node) is about ten times faster than that in different core-groups. On the other hand, amplitudes being stored in one SW26010 processor or two has almost no influence.

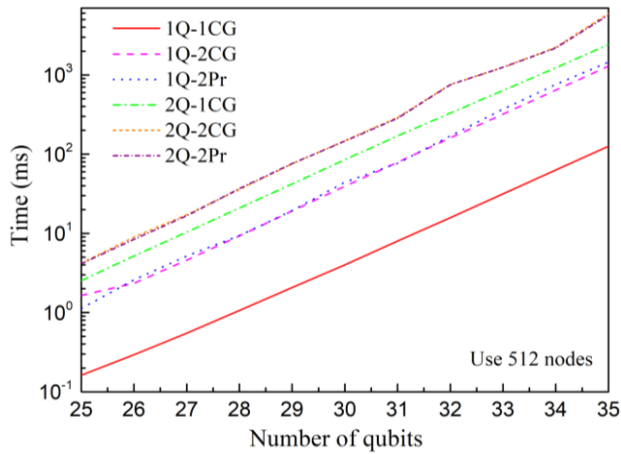

**Fig. 4** The time span of performing one- and two-qubit gates in different cases on the full amplitude mode. 1Q-1CG means that a one-qubit gate acts on a state stored in one core-group. 1Q-2CG means that a one-qubit gate acts on a state stored in two different core-groups, but in the same

SW26010 processor. 1Q-2Pr means that a one-qubit gate acts on a state stored in two SW26010 processors. 2Q represents the two-qubit gate

For the partial amplitude mode, we simulate a sequence of RQCs with 4096 nodes. The running time is shown in Fig. 5. In addition to the numbers of qubit and depth, the structure of the lattice of qubits also has a big impact on the running time, as shown by the results of 60 qubits (6×10 and 5×12).

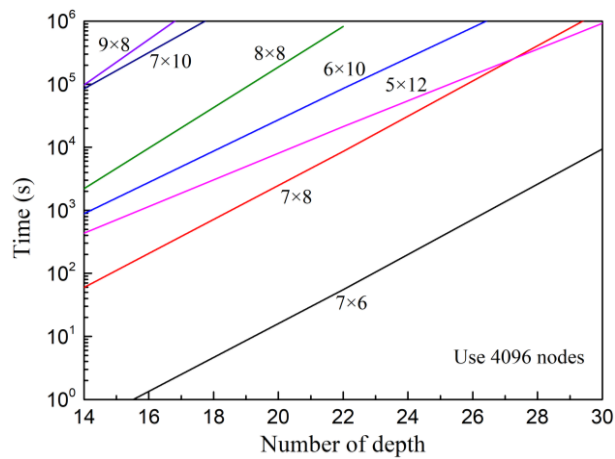

**Fig. 5** The time span of executing RQCs on the partial amplitude mode. The product of two numbers represents the total number of qubits

For the single amplitude mode, we simulate RQCs with 49, 110 and 200 qubits using 256 nodes. The running time is shown in Fig. 6. By taking advantage of the distributed computing system, we accomplished the simulation of circuits with up to 200 qubits and 21 depths.

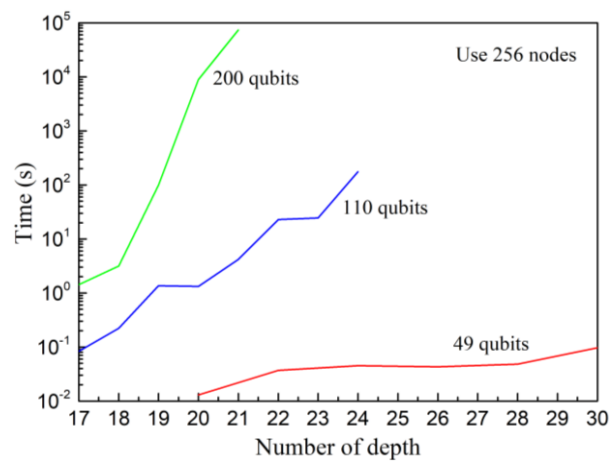

**Fig. 6** The time span of executing random quantum circuits on the single amplitude mode. The circuit with 200 qubits and 21 depths was simulated

3.2 Quantum fast Poisson solver
The Poisson equation is a widely used partial differential equation across many areas of physics and engineering. For instance, when simulating the dynamic process of ocean current, it is notoriously difficult to solve the Navier-Stokes equations directly [22]. The problem can be transformed to the solution of the Poisson equation using the vortex-in-cell model under certain conditions [23]. Therefore, solving the Poisson equation constitutes the most computationally intensive part of the ocean current simulation. We develop a quantum algorithm for solving the multi-dimensional Poisson equation. It could provide an exponential speedup to some degree over the classical counterparts [24]. Here, we remark that for the one-dimensional Poisson equation, there may exist more efficient quantum algorithms [25]. It could be implemented on the near-term NISQ devices. We leave this point to future work.

The general idea of our quantum fast Poisson solver is straightforward. First, we discretize the Laplacian operator to a square matrix using the central difference approximation, and then solve the resulting linear system of equations using the Harrow-Hassidim-Lloyd (HHL) algorithm [33].

The general circuit representation of our algorithm is shown in Fig. 7. It consists of three main stages, i.e., the phase estimation, controlled rotation and uncomputation. The phase estimation is to approximate the eigenvalues of the discretized matrix, and then entangle the states encoding the eigenvalues with the corresponding eigenstates. Hamiltonian simulation of $e^{iAt}$ is the crucial part of phase estimation, which use the techniques of sine transformation [34] and phase kickback [35]. After the phase estimation, the controlled rotation performs the linear map taking the state $|\lambda_j\rangle$ to $1/\lambda_j |\lambda_j\rangle$, where $\lambda_j$ is the eigenvalue. The rotation angles are prepared by evaluating an arc cotangent function by the ANGLE module in Fig. 7 [24]. Then, uncompute the stages of phase estimation and rotation-angle preparation to evolve the state of registers B, E and A back to the initial state. The computation process is finished by performing the measurement. If the result of the register Ancilla is $|1\rangle$, the solutions of the Poisson equation are encoded successfully in the final state.

The complexity of our algorithm is $O(d \log^2(\varepsilon^{-\alpha}))$ in qubits and $O(\kappa d \log^3(\varepsilon^{-\alpha}))$ in quantum operations, where $\varepsilon$ is the error of the solution, $d$ the dimension of Poisson equation, $\alpha > 0$ a smoothness constant and $\kappa$ the condition number of the discretized matrix [24]. On the other hand, any direct or iterative classical algorithms have a cost of at least $\varepsilon^{-\alpha d}$ [36]. Thus, our quantum Poisson solver could provide an exponential speedup over classical methods in the terms of dimension.

To demonstrate the correctness of the algorithm, we propose a simplified version of the circuit with four discretized points. The circuit consists of 38 qubits and 800 gates as shown in Fig. 8. It is simulated using the full amplitude mode, and the run time is 20 mins with 4096 nodes. The input state is $\frac{1}{\sqrt{2}}|01\rangle + \frac{1}{2}|10\rangle + \frac{1}{2}|11\rangle$. This corresponds to a Poisson equation with the solution of (0.9053, 1.1036 0.8018), which turns to (0.553 0.674 0.490) after normalization. The output state is $0.551|01\rangle + 0.675|10\rangle + 0.491|11\rangle$,

which is consistent with the real solution with an error less than 0.5%. The running results verify the correctness of our algorithm.

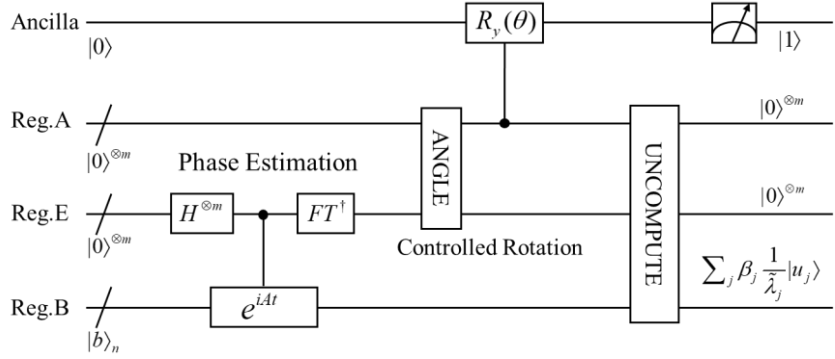

**Fig. 7** The overall quantum circuit of the quantum fast Poisson solver. It has four main registers, i.e. register B, E, A and Ancilla. Register B is used to encode the coefficients of Poisson equation. It is the input of the circuit. The approximated eigenvalues are stored in register E. Register A is used to store angles for the controlled rotation operation. The Ancilla register will transduce the reciprocals of eigenvalues to the amplitudes

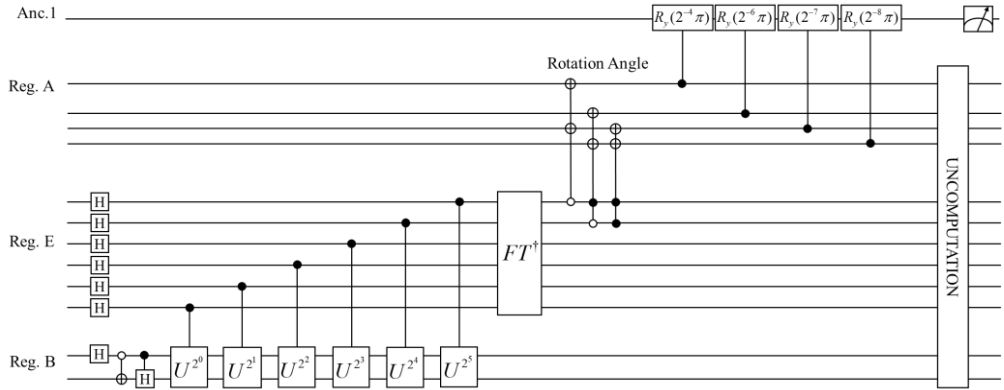

**Fig. 8** A simplified circuit of the quantum fast Poisson solver. It was executed successfully on the full amplitude mode. More information of this circuit can be found in Ref. [24]

3.3 Quantum arithmetic of transcendental functions

Quantum arithmetic in the computational basis constitutes the fundamental component of many circuit-based quantum algorithms. A vast amount of literature provided quantum circuits for solving the algebraic functions, including the addition [26], multiplication [27], reciprocal [28], and square root [29] operations, etc. However, studies about the higher-level transcendental functions are scare [30-31]. We develop a novel quantum algorithm, the qFBE (quantum Function-value Binary Expansion) method, to evaluate the transcendental functions [32]. The qFBE provides a unified and programmed solution for the evaluation of logarithmic, exponential, trigonometric and inverse trigonometric functions.

The qFBE algorithm belongs to a class of digit-recurrence method. It transforms the evaluation of transcendental functions to the recursive calculating of lower-level algebraic functions and outputting the binary string of function values digit-by-digit.

Such properties make the circuit design more compact and modular. Furthermore, it uses algebraic functions as subroutines, so the existing vast amount of quantum circuits for algebraic functions can be utilized directly to optimize the qFBE circuit.

The basic principle of the qFBE method can be well illustrated by the example of arctan(x) function,

$$\frac{\arctan(x)}{\pi} = \sum_{n \geq 0,\, a_n < 0} \frac{1}{2^{n+1}} \quad (for\ x > 0)$$

$$\text{with } a_0 = x,\ a_{n+1} = \frac{2a_n}{1 - a_n^2}\ (a_{n+1} = -\infty\ if\ a_n = \pm 1). \tag{6}$$

That is, the function value of arctan(x) can be expanded as a binary decimal, and each bit is determined based on the value of the iteration variable $a_n$. There exist similar recursions for other functions [37]. Formally, some applicable functions defined as $f: I \rightarrow [0,1]$, with $I \subseteq \mathbb{R}$ can be expanded as

$$f(x) = \sum_{n \geq 0,\, a_n \in D_1} \frac{1}{2^{n+1}}.$$

$$\text{with } a_0 = x,\ a_{n+1} = \begin{cases} r_0(a_n) & if\ a_n \in D_0 \\ r_1(a_n) & if\ a_n \in D_1 \end{cases}. \tag{7}$$

The $D_0$ and $D_1$ are subintervals of $I$ with $D_0 \cup D_1 = I$, $D_0 \cap D_1 = \emptyset$; $r_0$ and $r_1$ are functions defined as $r_0: D_0 \rightarrow I$, $r_1: D_1 \rightarrow I$. In fact, the challenge is to find proper $r_0(x)$ and $r_1(x)$ for a certain function $f$. This is the core idea of the qFBE method.

The logarithmic and inverse trigonometric functions can be evaluated using Eq. (7). They have simple $r_0(x)$ and $r_1(x)$, which are the lower-level algebraic functions. We list these functions as Group 1, i.e.,

**Group 1**: $\log_2(x)$, $\ln(x)$, $\arccos(x)$, $\arcsin(x)$, $\text{arccot}(x)$, $\arctan(x)$.

On the other hand, the inverse functions of Group 1, namely the exponential and trigonometric functions do not have proper $r_0(x)$ and $r_1(x)$. Therefore, they cannot be evaluated by the qFBE method. Fortunately, we find that they could be evaluated by inverting the qFBE procedure as follows,

$$f(x) = a_{n+1}$$
$$\text{with } x = (0.v_{n-1}v_{n-2}\cdots v_1 v_0),\ v_i \in \{0,1\}, \tag{8}$$
$$\text{and } a_0 = const,\ a_{i+1} = \begin{cases} r_0^{-1}(a_i) & if\ v_i = 0 \\ r_1^{-1}(a_i) & if\ v_i = 1 \end{cases}.$$

Apparently, qFBE [see Eq. (7)] outputs the function value digit-by-digit in a recursive way, while the inverse qFBE [see Eq. (8)] approximates the function value step-by-step in an iterative way. We list these functions as Group 2, i.e.,

**Group 2**: $2^x$, $e^x$, $\cos(x)$, $\sin(x)$, $\cot(x)$, $\tan(x)$.

The recursive property of the qFBE makes the quantum circuit design more compact and modular. The general structure of the quantum circuits for evaluating functions in Groups 1 and 2 are shown in Fig. 9 (a) and (b), respectively. For the qFBE method, the circuit consists of ($n$-1) modules, which actually implement the recursions in Eq. (7).

Each module outputs one bit of the solution. For the inverse qFBE method, the circuit consists of $n$ modules, which approximate the function value step-by-step according to Eq. (8). The last module outputs the final solution.

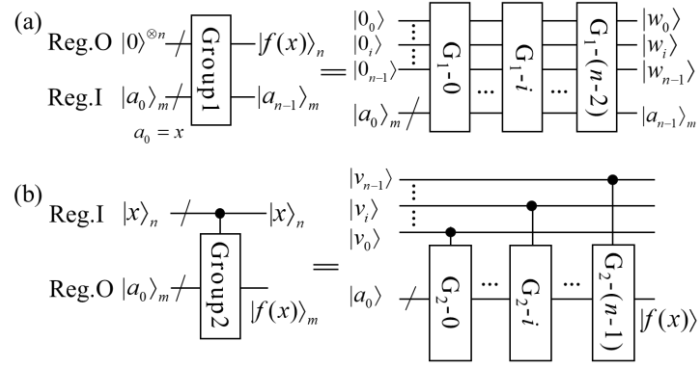

**Fig. 9** The overall circuits for evaluating functions in Group 1 (a), and Group 2 (b). The Group 1 circuit consists of $n$-1 modules $G_1$-$i$, which implement the recursions in Eq. (7). Similarly, the Group 2 circuit consists of $n$ modules $G_2$-$i$, which corresponds to the iterations in Eq. (8). Both circuits mainly include two registers for inputs and outputs

The complexity of evaluating transcendental functions by the qFBE method is $nO(m)$ in qubits and $nO(m^2)$ in quantum gates, where $n$ is the number of qubits to encode input or output and $m$ the number of qubits to store the intermediate values. The cost of our method is comparable with the best known results [31] at worst case; while when the input binary has a small number of bits, our method cost much lower. Furthermore, all digits of the binary output can be exact, which makes the control of error propagation easy. The qFBE method provides a unified and programmed solution for most transcendental functions, and the circuits are compact and modular which are easy to be implemented on the virtual or the future real quantum machine.

We present the complete quantum circuits for all the functions in Groups 1 and 2 and demonstrate the correctness of these circuits on the simulator. The configurations when implementing the circuits are listed in Tab. 1. The ARCCOT, COS, ARCCOS, COT, EXP and LOG represent the functions of $\mathrm{arccot}(x)/\pi$, $\cos(\pi x)$, $\mathrm{arccos}(x)/\pi$, $\cot(\pi x)$, $2^x$ and $\log_2 x$, respectively. The values of qubits, gates, nodes and run time in the table just provide a description about the circuits executed on the simulator. They have nothing to do with the cost and efficiency of the qFBE algorithm. The inputs and outputs of these circuits are easy to understand. Taking COS as an example, the input states are $|.00\rangle, |.01\rangle, |.10\rangle, |.11\rangle$, and the corresponding outputs are $|01.000\rangle, |00.101\rangle, |00.000\rangle, |11.011\rangle$, respectively. The running results verify the correctness of our algorithm.

Table 1 The configurations of implementing the circuits on the simulator

|  | ARCCOT | COS | ARCCOS | COT | EXP | LOG |
|---|---|---|---|---|---|---|
| Qubits | 24 | 29 | 30 | 30 | 32 | 40 |
| Basic Gates [a] | 1000 | 730 | 770 | 1100 | 820 | 1260 |
| Nodes | 4 | 8 | 16 | 16 | 64 | 16384 |
| Run time (min) | 4 | 14 | 13 | 21 | 14 | 23 |

[a] They include the *H*, *X*, *C-NOT*, *SWAP* and *TOFFOLI* gates as shown in Appendix A.

**4 Conclusions**

We have developed an efficient quantum circuit simulator on the Sunway TaihuLight supercomputer. The simulator possesses three working modes, being capable of calculating the full, partial and single amplitudes of a quantum state. The three modes are built using entirely different methodologies. They are the direct evolution of quantum states, circuit partition by decomposing controlled-Z gate and the complex undirected graphical model. Our simulator has the function of emulating the effects of noise, and it supports many kinds of useful quantum gates and operations. To make full use of the Sunway distributed system, the simulation was implemented in a two-level parallel way. With 16384 computational nodes, roughly 10% of the computing resource of the Sunway, random quantum circuits with up to 40, 75 and 200 qubits can be simulated on full, partial and single amplitude modes, respectively.

Based on the simulator, we further developed the quantum algorithms for solving the Poisson equations and for quantum arithmetic of evaluating transcendental functions. The present quantum fast Poisson solver takes the HHL algorithm as the framework, and provides an exponential speedup over the classical methods in the terms of dimension. The qFBE method provides a unified and programmed way of evaluating the transcendental functions, including the logarithmic, exponential, arc-cosine, arc-sine, cosine, sine, arc-cotangent, arc-tangent, cotangent and tangent functions.

For future work, we will (1) advance the study of quantum Poisson solver to further reduce the algorithm complexity and quantify the effect of noise, and (2) optimize the qFBE circuits by selecting the proper circuits of evaluating algebraic functions. Furthermore, we will expand the applications of the present simulator to other fields, like variational quantum algorithms and quantum machine learning.

**Acknowledgements**

We are very grateful to the National Supercomputing Center in Wuxi for the great computing resource. The present work is financially supported by the National Natural Science Foundation of China (Grant No. 61575180, 61701464, 11475160) and the Pilot National Laboratory for Marine Science and Technology (Qingdao).

**Author contributions**

Z. Wang designed the quantum algorithms for applications, participated partially in the design of the simulator and prepared the manuscript. Z. Chen wrote the simulation programs and tested the simulator. S. Wang and W. Li designed the quantum algorithms for applications and tested the simulator. Y. Gu, G. Guo and Z. Wei planned, organized and supervised the project. All authors discussed the results and reviewed the manuscript.

**Appendix A**

In this supplementary material, a description of the instruction set of our simulator is given in Section A-1. Section A-2 shows an example of the input and output to illustrate the way of running the simulator.

**A-1 Instruction set**
**QINIT**
Description: specify the number of qubits of the circuit which are initialized as |0>.
Syntax: QINIT $n$
Argument: $n$ is the number of qubits of the quantum circuit.

**CREG**
Description: allocate classical registers for storing the measurement results.
Syntax: CREG $m$
Argument: $m$ is the number of classical registers.

**H/X/Y/Z/S/T gate**
Description: perform Hadamard/Pauli-X/Pauli-Y/Pauli-Z/phase/T operations.
Syntax: H/X/Y/Z/S/T $i$
Argument: $i$ represents a certain qubit ranging from 0 to $n$-1.

Operation: $H = \frac{\sqrt{2}}{2}\begin{bmatrix} 1 & 1 \\ 1 & -1 \end{bmatrix}, X = \begin{bmatrix} 0 & 1 \\ 1 & 0 \end{bmatrix}, Y = \begin{bmatrix} 0 & -i \\ i & 0 \end{bmatrix}, Z = \begin{bmatrix} 1 & 0 \\ 0 & -1 \end{bmatrix}, S = \begin{bmatrix} 1 & 0 \\ 0 & i \end{bmatrix}, T = \begin{bmatrix} 1 & 0 \\ 0 & e^{i\pi/4} \end{bmatrix}.$

**RX/RY/RZ gates**
Description: rotate the qubit about x-/y-/z-axis.
Syntax: RX/RY/ RZ $i$, "$\theta$"
Argument: $i$ represents a certain qubit from 0 to $n$-1 and $\theta$ is the rotation angle in radian.

Operation: $RX(\theta) = \begin{bmatrix} \cos\frac{\theta}{2} & -i\sin\frac{\theta}{2} \\ -i\sin\frac{\theta}{2} & \cos\frac{\theta}{2} \end{bmatrix}, RY(\theta) = \begin{bmatrix} \cos\frac{\theta}{2} & -\sin\frac{\theta}{2} \\ \sin\frac{\theta}{2} & \cos\frac{\theta}{2} \end{bmatrix}, RZ(\theta) = \begin{bmatrix} e^{-i\theta/2} & 0 \\ 0 & e^{i\theta/2} \end{bmatrix}.$

**U4 gates**
Description: perform arbitrary sing-qubit gate.
Syntax: U4 $i$, "$u_0, u_1, u_2, u_3$"
       U4 $i$, "$\alpha, \beta, \gamma, \delta$"
Argument: $i$ represents a certain qubit from 0 to $n$-1. The $u_0, u_1, u_2, u_3$ or $\alpha, \beta, \gamma, \delta$ are the elements or angles in radian in the matrix of the operator.

Operation: $U4 = \begin{bmatrix} u_0 & u_1 \\ u_2 & u_3 \end{bmatrix}, \quad U4 = \begin{bmatrix} e^{i(\alpha-\frac{\beta}{2}-\frac{\delta}{2})}\cos\frac{\gamma}{2} & -e^{i(\alpha-\frac{\beta}{2}+\frac{\delta}{2})}\sin\frac{\gamma}{2} \\ e^{i(\alpha+\frac{\beta}{2}-\frac{\delta}{2})}\sin\frac{\gamma}{2} & e^{i(\alpha+\frac{\beta}{2}+\frac{\delta}{2})}\cos\frac{\gamma}{2} \end{bmatrix}.$

**CNOT/CZ/CR gate**
Description: perform Pauli-X/Pauli-Z/Rotation-about-z-axis on the target qubit when the control qubit is |1>.

Syntax: CNOT/CZ $i, j$
      CR $i, j,$ "$\theta$"

Argument: $i$ and $j$ represent two different qubits ranging from 0 to $n$-1, where $i$ is the control qubit and $j$ the target qubit. The $\theta$ is rotation angle in radian.

Operation: $CNOT = \begin{bmatrix} 1 & 0 & 0 & 0 \\ 0 & 1 & 0 & 0 \\ 0 & 0 & 0 & 1 \\ 0 & 0 & 1 & 0 \end{bmatrix}$, $CZ = \begin{bmatrix} 1 & 0 & 0 & 0 \\ 0 & 1 & 0 & 0 \\ 0 & 0 & 1 & 0 \\ 0 & 0 & 0 & -1 \end{bmatrix}$, $CR = \begin{bmatrix} 1 & 0 & 0 & 0 \\ 0 & 1 & 0 & 0 \\ 0 & 0 & 1 & 0 \\ 0 & 0 & 0 & e^{i\theta} \end{bmatrix}$.

**SWAP/iSWAP gate**

Description: swap the states of two qubits/swap the states of two qubits and add a $\pi/2$ phase.

Syntax: SWAP/ISWAP $i, j$

Argument: $i$ and $j$ represent two different qubits ranging from 0 to $n$-1.

Operation: $SWAP = \begin{bmatrix} 1 & 0 & 0 & 0 \\ 0 & 0 & 1 & 0 \\ 0 & 1 & 0 & 0 \\ 0 & 0 & 0 & 1 \end{bmatrix}$, $iSWAP = \begin{bmatrix} 1 & 0 & 0 & 0 \\ 0 & 0 & -i & 0 \\ 0 & -i & 0 & 0 \\ 0 & 0 & 0 & 1 \end{bmatrix}$.

**TOFFOLI gate**

Description: perform NOT operation on the target qubit when the two control qubits are all in |1>.

Syntax: TOFFOLI $i, j, k$

Argument: $i, j$ and $k$ represent three different qubits ranging from 0 to $n$-1, where $i, j$ are the control qubits and $k$ the target qubit.

Operation: $TOFFOLI = \begin{bmatrix} 1 & 0 & 0 & 0 & 0 & 0 & 0 & 0 \\ 0 & 1 & 0 & 0 & 0 & 0 & 0 & 0 \\ 0 & 0 & 1 & 0 & 0 & 0 & 0 & 0 \\ 0 & 0 & 0 & 1 & 0 & 0 & 0 & 0 \\ 0 & 0 & 0 & 0 & 1 & 0 & 0 & 0 \\ 0 & 0 & 0 & 0 & 0 & 1 & 0 & 0 \\ 0 & 0 & 0 & 0 & 0 & 0 & 0 & 1 \\ 0 & 0 & 0 & 0 & 0 & 0 & 1 & 0 \end{bmatrix}$.

**DAGGER & ENDDAGGER**

Description: perform the inverse operation of a group of gates in the circuit.

Syntax: DAGGER
    $M$
    ENDDAGGER

Argument: DAGGER is the starting mark and ENDDAGGER is the terminating mark.

Between them, the group of gates *M* is converted to its transpose conjugate

$M^{\dagger}$ to be implemented. This instruction could be used in the nested way.

**CONTROL & ENDCONTROL**
Description: perform the control operation for a group of gates in the circuit.
Syntax: CONTROL *i*
     *M*
     ENDCONTROL *i*
Argument: CONTROL is the starting mark and END CONTROL is the terminating mark. Between them, the group of gates *M* is implemented controlled by the qubit *i*. This instruction could be used in the nested way.

**MEASURE**
Description: perform partial measurement on certain qubits and normalize the left quantum states to move on evolving.
Syntax: MEASURE *i*, $*j*
Argument: *i* represents a certain qubit ranging from 0 to *n*-1, and *j* behind $ represents a classical register specified by CREG. The measurement results are stored in the classical registers.

**PMEASURE**
Description: calculate the probability of the states in the space spanned by the specified qubits. No quantum states change in this measurement.
Syntax: PMEASURE *i*, *j*, …, *k*
Argument: *i* to *k* represent the qubits to measure which range from 0 to *n*-1. The sum of the probability of all the measured states should equal to 1.

**A-2 Input and output**
    An example of input is shown below with the instructions defined in Section S1. For simplicity, the input is shown in three columns. Generally, the input consists of three parts: first the configurations of qubits and classical registers, then the body of quantum operations, finally the measurement.
    Upload the input script and submit the task to the supercomputer to start to run. If the input is written legally, measurement results will be stored in the output file; otherwise, the error information will be stored in the log.txt file. An example of output is shown below. The first column is the computational basis to measure, and the second one is the probability of the corresponding basis (corresponding to the PMEASURE instruction in the input).

| Example: Input | | | Example: output |
|---|---|---|---|
| %Configure | S 2 | X 2 | 000: 0.820082 |
| QINIT 5 | H 0 | H 2 | 001: 0 |
| CREG 3 | SWAP 0,2 | CR 1,2,"pi/2" | 010: 0.106694 |
| %Operate | TOFFOLI 0,1,3 | CR 0,2, "pi/4" | 011: 0 |
| H 1 | RY 3,"-pi/4" | H 1 | 100: 0.0549175 |
| S 2 | TOFFOLI 0,1,3 | CR 0,1,"pi/2" | 101: 0 |
| X 2 | RY 3,"pi/4" | H 0 | 110: 0.0183058 |
| H 2 | CNOT 0,4 | SWAP 0,2 | 111: 0 |
| CR 1,2,"pi/2" | RY 4,"-pi/8" | ENDDAGGER | |
| CR 0,2, "pi/4" | CNOT 0,4 | %Measure | |
| H 1 | RY 4,"pi/8" | PMEASURE 4,3,0 | |
| CR 0,1,"pi/2" | DAGGER | | |